# A Comparative Study of Replication Techniques in Grid Computing Systems

Sheida Dayyani
Department of Computer Engineering
Sheikh Bahaei University
Isfahan, Iran
sh.dayani@shbu.ac.ir

Mohammad Reza Khayyambashi
Department of Computer Engineering
University of Isfahan
Isfahan, Iran
m.r.khayyambashi@eng.ui.ac.ir

*Abstract*—Grid Computing is a type of parallel and distributed systems that is designed to provide reliable access to data and computational resources in wide area networks. These resources are distributed in different geographical locations, however are organized to provide an integrated service. Effective data management in today`s enterprise environment is an important issue. Also, Performance is one of the challenges of using these environments. For improving the performance of file access and easing the sharing amongst distributed systems, replication techniques are used. Data replication is a common method used in distributed environments, where essential data is stored in multiple locations, so that a user can access the data from a site in his area. In this paper, we present a survey on basic and new replication techniques that have been proposed by other researchers. After that, we have a full comparative study on these replication strategies. Also, at the end of the paper, we summarize the results and points of these replication techniques.

## I. INTRODUCTION

Computing infrastructure and network application technologies have come a long way over the past years and have become more and more detached from the underlying hardware platform on which they run. At the same time computing technologies have evolved from monolithic to open and then to distributed systems [1].

Nowadays, there is a tendency of storing, retrieving, and managing different types of data such as experimental data that are produced from many projects [2]. This data plays a fundamental role in all kinds of scientific applications such as particle physics, high energy physics, data mining, climate modeling, earthquake engineering and astronomy, to cite a few, manage and generate an important amount of data which can reach terabytes and even petabytes, which need to be shared and analyzed [3], [4], [5].

Storing such amount of data in the same location is difficult, even impossible. Moreover, an application may need data produced by another geographically remote application. For this reason, a grid is a large scale resource sharing and problem solving mechanism in virtual organizations and is suitable for the above situation [6], [7], [8]. In addition, users can access important data that is available only in several locations, without the overheads of replicating them locally. These services are provided by an integrated grid service platform so that the user can access the resource transparently and effectively [2], [6]. Managing this data in a centralized location increases the data access time and hence much time is taken to execute the job. So to reduce the data access time, "Replication" is used [3], [4].

The replication is the process of creation and placement of the copies of entities software. The phase of creation consists in reproducing the structure and the state of the replicated entities, whereas the phase of placement consists in choosing the suitable slot of this new duplication, according to the objectives of the replication. So, replication strategy can shorten the time of fetching the files by creating many replicas stored in appropriate locations [9], [10]. By storing the data at more than one site, if a data site fails, a system can operate using replicated data, thus increasing availability and fault tolerance. At the same time, as the data is stored at multiple sites, the request can find the data close to the site where the request originated, thus increasing the performance of the system. But the benefits of replication, of course, do not come without overheads of creating, maintaining and updating the replicas [11].

There is a fair amount of work on data replication in grid environments. Most of the existing work focused on mechanisms for create, decision and delete replicas. The purpose of this document is to review various replication techniques and compare these techniques which have been presented by other researches in different distributed architectures and topologies.

The rest of this paper is organized as follows. In the second section, we present an overview of grid systems, types of grids and topologies that exist for grid systems. The third section describes replication scenario, challenges and parameters of evaluating replication techniques. Section four takes a closer look on basic and new existing data replication strategies in grid environment. In section five, we present a comparative study on the replication techniques that were discussed in the previous Section. Finally, section six will be reserved for the conclusion and a summary of discussed replication techniques results.

## II. GRID SYSTEMS

A large number of scientific and engineering applications require a huge amount of computing time to carry out their experiments by simulation. Research driven by this has promoted the exploration of a new architecture known as "The Grid" for high performance distributed application and systems [12]. In [13], Foster defines the Grid concept as "coordinated resource sharing and problem solving in dynamic, multi-institutional virtual organizations". There are different types and topologies of Grid developed to emphasize special functions that will be defined in the two next sections.

### A. Types of Grid

Grid computing can be used in a variety of ways to address various kinds of application requirements and it has three primary types. Of course, there are no hard boundaries between these grid types and often grids may be a combination of two or more of these [14]. Types of grids are summarized below:

- **Computational grid:** Computational grid is focused on setting aside resources specifically for computing power. Such as most of the machines are high-performance servers [14].
- **Scavenging grid:** Scavenging grid is most commonly used with large numbers of desktop machines that are scavenged for available CPU cycles and other resources. Owners of the desktop machines are usually given control over when their resources are available to participate in the grid [14].
- **Data grid:** Data grid is a collection of geographically distributed computer resources that these resources may be located in different parts of a country or even in different countries [10]. For example, you may have two universities doing life science research, each with unique data. A grid connects all these locations and enables them to share their data, manage the data, and manage security issues such as who has access to which data [15], [16].

### B. Grid Topologies

In this section we present an overview of major grid topologies. The performance of replication strategies is highly dependent on the underlying architecture of grid [17], [18].

Hierarchical and tree models are used where there is a single source for data and the data has to be distributed among collaborations worldwide [17], [18]. The Figure 1 and Figure 2, shows the hierarchical and tree models respectively.

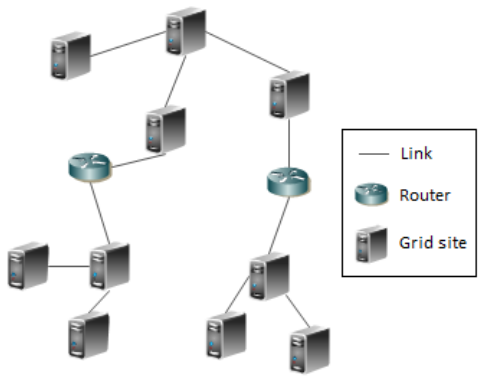

Figure 1. An example of Hierarchical topology.

A tree topology also has shortcomings. The tree structure of the grid means that there are specific paths to the messages and files can travel to get to the destination. Furthermore, data transference is not possible among sibling nodes or nodes situated on the same tier [17], [18].

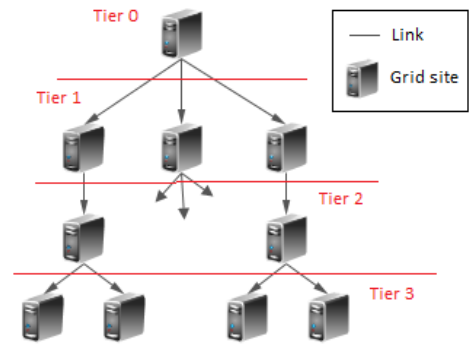

Figure 2. An example of Tree topology.

Peer to Peer (P2P) systems overcome these limitations and offer flexibility in communication among components. A P2P system is characterized by the applications that employ distributed resources to perform functions in a decentralized manner. From the viewpoint of resource sharing, a P2P system overlaps a grid system. The key characteristic that distinguishes a P2P system from other resource sharing systems is its symmetric communication model between peers, each of which acts as both a server and a client [17], [18]. The Figure 3, shows an example of the P2P structure.

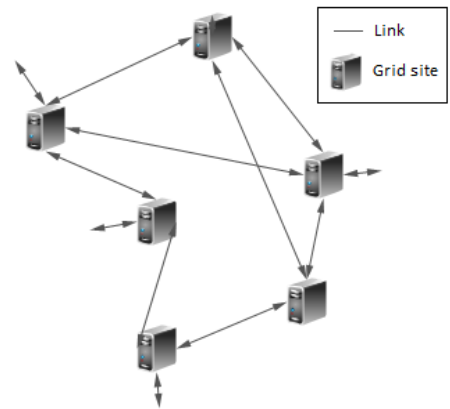

Figure 3. An example of Peer to Peer topology.

Hybrid Topology is simply a configuration that contains an architecture consisting of any combination of the previous mentioned topologies. It is used mostly in situations where researches working on projects want to share their results to further research by making it readily available for collaboration [17], [18]. A hybrid model of a hierarchical grid with peer linkages at the edges is shown in Figure 4.

A hybrid topology can carry features of both tree and P2P architectures and thus can be used for better performance of a replication strategy [15].

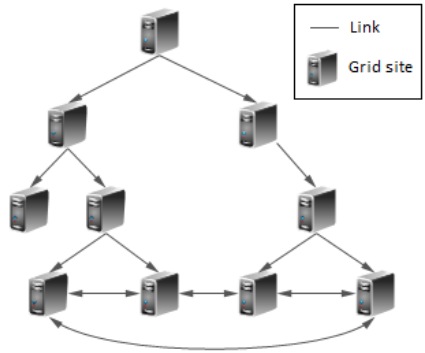

Figure 4. An example of Hybrid topology .

## III. DATA MANAGEMENT IN GRIDS

An important technique for data management in grid systems is the replication technique. Data replication is characterized as an important optimization technique in Grid for promoting high data availability, low bandwidth consumption, increased fault tolerance, and improved scalability. The goals of replica optimization is to minimize file access times by pointing access requests to appropriate replicas and pro-actively replicating frequently used files based on access statistics gathered.

Generally, replication mechanism determines which files should be replicated, when the new replicas should be created and where the new replicas should be placed [4], [9], [15]. In the rest of this section, we discuss about data replication scenario, challenges and parameters.

### A. Data Replication Scenario

The main aims of using replication are to reduce access latency and bandwidth consumption. The other advantages of replication are that it helps in load balancing and improves reliability by creating multiple copies of the same data [4], [15].

Replication schemes can be classified as static and dynamic. In **static replication**, a replica persists until it is deleted by users or its duration is expired. The drawback of static replication is evident when client access patterns change greatly in the Data. Static replication can be used to achieve some of the above mentioned goals but the drawback with static replication is that it cannot adapt to changes in user behavior. The replicas have to be manually created and managed if one were to use static replication. But, in **dynamic replication**, replica creation, deletion and management are done automatically. Dynamic strategies have the ability to adapt to changes in user behavior [19].

Various combinations of events and access scenarios of data are possible in a distributed replicated environment. The three fundamental questions any replica placement strategy has to answer are as follow that Depending on the answers, different replication strategies are born [4], [15]:
- When the replicas should be created?
- Which files should be replicated?
- Where the replicas should be placed?

### B. Data Replication Challenges

Using replication strategies in grid environment may cause some challenges. The four important challenges in replicated environments are as follow [11]:
- Time of creation of a new replica: If strict data consistency is to be maintained, performance is severely affected if a new replica is to be created. As sites will not be able to fulfill request due to consistency requirements.
- Data Consistency: Maintaining data integrity and consistency in a replicated environment is of prime importance. High precision applications may require strict consistency of the updates made by transactions.
- Lower write performance: Performance of write operations can be dramatically lowered in applications requiring high updates in replicated environment, because the transaction may need to update multiple copies.
- Overhead of maintenance: If the files are replicated at more than one site, it occupies storage space and it has to be administered. Thus, there are overheads in storing multiple files.

### C. Data Replication Evaluation

Almost all the replications strategies try to reduce the **access latency** thus reducing the job response time and hence increase the performance of the grids. Similarly almost all the replication strategies try to reduce the **bandwidth consumption** to improve the availability of data and performance of the system. The target is to keep the data as close to the user as possible, so that data can be accessed efficiently. Some of the replication strategies explicitly target to provide a **balanced workload** on all the data servers. This helps in increasing the performance of the system and provides better response time. With more number of replicas in a system the cost of maintaining them becomes an overhead for the system. Some of the strategies aim to make only an optimal number of replicas in the data grid. This ensures that the storage is utilized in an optimal way and the **maintenance cost** of replica is minimized. Some strategies target the strategic placement of the replicas along with an optimal number of replicas. The **strategic placement** of replicas is a very important factor because it is integrated with few other very important factors. For example, if the replicas are placed on the optimal locations it helps to optimize the workload of different servers. It is also related with the cost of the maintenance. If a strategy goes on replicating a popular file blindly, it will create too many replicas thus increasing the burden for the system as replica maintenance costs will become too high [20].

**Job execution time** is another very important parameter. Some replication strategies target to minimize the job execution time with optimal replica placement. The idea is to place the replicas closer to the users in order to minimize the response time, and thus the job execution time. This will increase the throughput of the system [20]. Only a few replication strategies have considered replication as an option to provide **fault tolerance** and **quality assurance**. All replication strategies use subset of these parameters [20].

## IV. REPLICATION TECHNIQUES

The role of a replication strategy is to identify when a replica should be created, where to place replicas, when to remove replicas and how to locate the best replica [21].

Several replication replacement strategies have been proposed in the past and they are the basics of other replication algorithms. Details of some important basic and new replication algorithms are as follows:

- **NO Replication strategy** will not create replica and therefore, the files are always accessed remotely. One example of the implemented strategy is the SimpleOptimizer algorithm [22], which never performs replication; rather it reads the required replica remotely. SimpleOptimizer algorithm is simple to implement and performs the best relative to other algorithms in terms of the storage space usage, but performs the worst in terms of job execution time and network usage [15].

- **Best client** creates replica at the client that has generated the most requests for a file, this client is called the best client. At a given time interval, each node checks to see if the number of requests for any of its file has exceeded a threshold, then the best client for that file is identified [15].

- **Cascading Replication** supports tree architecture. The data files generated in the top level and once the number of accesses for the file exceeds the threshold, then a replica is created at the next level, but on the path to the best client, and so on for all levels, until it reaches to the best client itself [15].

- **Plain Cashing:** The client that requests a file stores a copy locally. If these files are large and a client has enough space to store only one file at a time, then files get replaced quickly [15].

- **Cashing plus Cascading** combines cascading and plain cashing strategies. The client caches file locally, and the server periodically identifies the popular files and propagates them down the hierarchy. Note that the clients are always located at the leaves of the tree but any node in the hierarchy can be a server. Specifically, a Client can act as a Server to its siblings. Siblings are nodes that have the same parent [15].

- **Fast Spread:** In this method a replica of the file is stored at each node along its path to the client. When a client requests a file, a copy is stored at each tier on the way. This leads to a faster spread of data. When a node does not have enough space for a new replica it deletes the least popular file that had come in the earliest [15].

- **Least Frequently Used (LFU)** strategy always replicates files to local storage systems. If the local storage space is full, the replica that has been accessed the fewest times is removed and then releases the space for new replica. Thus, LFU deletes the replica which has less demand (less popularity) from the local storage even if the replica is newly stored [23].

- **Least Recently Used (LRU)** strategy always replicates files to local storage system. In LRU strategy, the requested site caches the required replicas, and if the local storage is full, the oldest replica in the local storage is deleted in order to free the storage. However, if the oldest replica size is less than the new replica, the second oldest file is deleted and so on [23].

- **Proportional Share Replica (PSR)** policy is an improvement in Cascading technique. The method is a heuristic one that places replicas on the optimal locations by assuming that the numbers of sites and the total replicas to be distributed are already known. Firstly an ideal load distribution is calculated and then replicas are placed on candidate sites that can service replica requests slightly greater than or equal to that ideal load [24].

- **Bandwidth Hierarchy Replication (BHR)** is a novel dynamic replication strategy which reduces data access time by avoiding network congestions in a data grid network. With BHR strategy, we can take benefits from "network-level locality" which represents that required file is located in the site which has broad bandwidth to the site of job execution. BHR strategy was evaluated by implementing in OptorSim simulator and the results show that BHR strategy can outperform other optimization techniques in terms of data access time when hierarchy of bandwidth appears in Internet. BHR extends current site-level replica optimization study to the network-level [25].

- **Simple Bottom-Up (SBU)** and **Aggregate Bottom-Up (ABU)** are two dynamic replication mechanisms that are proposed in the multi-tier architecture for data grids. The SBU algorithm replicates the data file that exceeds a pre-defined threshold for clients. The main shortcoming of SBU is the lack of consideration to the relationship with historical access records. For the sake of addressing the problem, ABU is designed to aggregate the historical records to the upper tier until it reaches the root. The results shown improvements against Fast Spread strategy. The values for interval checking and threshold were based on data access arrival rate, data access distribution and capacity of the replica servers [16].

- **Multi-objective approach** is a method exploiting operations research techniques that is proposed for replica placement. In this method, replica placement decision is made considering both the current network status and data request pattern. The problem is formulated in p-median and p-center models to find the p replica placement sites. The p-center problem targets to minimize the max response time between user site and replica server whereas the p-median model focuses on minimizing the total response time between the requesting sites and the replication sites [26], [27].

- **Weight-based dynamic replica replacement** strategy calculates the weight of replica based on the access time in the future time window on the last access history. After that, calculate the access cost which embodies the number of replicas and the current bandwidth of the network. The replicas with high weight will be helpful to improve the efficiency of data access, so they should be retained and then the replica with low weight will not make sense to the rise of data access efficiency, and therefore, should be deleted. The access history defines based on the zipf-like distribution [28].

- **Latest Access Largest Weight (LALW)** is a dynamic data replication mechanism. LALW selects a popular file for replication and calculates a suitable number of copies and grid sites for replication. By associating a different weight to each historical data access record, the importance of each record is differentiated. A more recent data access record has a larger weight. It indicates that the record is more pertinent to the current situation of data access [29].

- **Agent-based replica placement algorithm** is proposed to determine the candidate site for the placement of replica. For each site that holds the master copies of the shared data files will deploy an agent. The main objective of an agent is to select a candidate site for the placement of a replica that reduces the access cost, network traffic and aggregated response time for the applications. Furthermore, in creating the replica an agent prioritizes the resources in the grid based on the resource configuration, bandwidth in the network and insists for the replica at their sites and then creates a replica at suitable resource locations [7].

- **Adaptive Popularity Based Replica Placement (APBRP)** is a dynamic replica placement algorithm, for hierarchical data grids which is guided by "file popularity". The goal of this strategy is to place replicas close to clients to reduce data access time while still using network and storage resources efficiently. The effectiveness of APBRP depends on the selection of a threshold value related to file popularity. APBRP determines this threshold dynamically based on data request arrival rates [30].

- **Efficient Replication strategy** is a replication strategy for dynamic data grids, which take into account the dynamic of sites. This strategy can increase the file availability, improved the response time and can reduce the bandwidth consumption. Moreover, it exploits the replicas placement and file requests in order to converge towards a global balancing of the grid load. This strategy will focus on read-only-access as most grids have very few dynamic updates because they tend to use a "load" rather than "update" strategy.

There are three steps provided by this algorithm, which are:
1. Selection of the best candidate files for replication; Selected based on requests number and copies number of each files.
2. Determination of the best sites for files placement which are selected in the previous step; Selected based on requests number and utility of each site regarding to the grid.
3. Selection of the best replica; Taking account the bandwidth and the utility of each site [31].

- **Value-based replication strategy (VBRS)** is proposed to decrease the network latency and meanwhile to improve the performance of the whole system. In VBRS, threshold was made to decide whether to copy the requested file, and then solve the replica replacement problem. VBRS has two steps; At the first steps, the threshold will be calculated to decide whether the requested file should be copied in the local storage site. Then at the second stage, the replacement algorithm will be triggered when the requested file needs to be copied at the local storage site does not have enough space. The replica replacement policy is developed by considering the replica's value which is based on the file's access frequency and access time. The experiment results show that the effectiveness of VBRS algorithm can reduce network latency [32].

- **Enhance Fast Spread (EPS)** is an enhanced version of Fast Spread for replication strategy in the data grid. This strategy was proposed to improve the total of response time and total bandwidth consumption. Its takes into account some criteria such as the number and frequency of requests, the size of the replica and the last time the replica was requested. EFS strategy keeps only the important replicas while the other less important replicas are replaced with more important replicas. This is achieved by using a dynamic threshold that determines if the requested replica should be stored at each node along its path to the requester [33].

- **Predictive hierarchical fast spread (PHFS)** is a dynamic replication method in multi-tier data grid environments which is an improve version of common fast spread. The PHFS tries to forecast future needs and pre-replicates the min hierarchal manner to increase locality in accesses and improve performance that consider spatial locality. This method is able to optimize the usage of storage resources, which not only replicates data objects hierarchically in different layers of the multi-tier data grid for obtaining more localities in accesses. It is a method intended for read intensive data grids. The PHFS method use priority mechanism and replication configuration change component to adapt the replication configuration dynamically with the obtainable condition. Besides that, it is developed on the basis of the concept that users who work on the same context will request some files with high probability [34].

- **Dynamic Hierarchical Replication (DHR)** is a dynamic replication algorithm for hierarchical structure that places replicas in appropriate sites. Best site has the highest number of access for that particular replica. This algorithm minimizes access latency by selecting the best replica when various sites hold replicas. The replica selection strategy of DHR algorithm, selects the best replica location for the users running jobs by considering the replica requests that waiting in the queue and data transfer time. It stores the replica in the best site where the file has been accessed most, instead of storing files in many sites [35].

- **Modified Latest Access Largest Weight (MLALW)** is a dynamic data replication strategy. This strategy is an enhanced version of Latest Access Largest Weight strategy. MLALW deletes files by considering three important factors:
1. Least frequently used replicas
2. Least recently used replicas
3. The size of the replica

MLALW stores each replica in an appropriate site in the region that has the highest number of access in future for that particular replica. The experiment results show that MLALW strategy gives a better performance compared to the other algorithms and prevents unnecessary creation of replica which leads to efficient storage usage [36].

## V. COMPARATIVE STUDY

In this section, we present a full comparative study on the replication techniques that were discussed in the previous section.

These twenty two replication strategies are compared in the Table 1, Table 2 and Table 3.

TABLE I. COMPARATIVE STUDY ON REPLICATION TECHNIQUES (A)

| Replication technique | Method | Performance metric | Topology | Scalability | Used storage | Simulator | Year | Additional feature |
|---|---|---|---|---|---|---|---|---|
| **Best Client [15]** | Replicates file to site that generates maximum number of requests | Response time, Bandwidth conservation | Tree structure (top-down) | Medium | Low | A grid simulator using PARSEC | 2001 | Need to compute number of request for each file |
| **Cascading [15]** | If number of requests exceeds threshold then replica trickles down to lower tier | Response time, Bandwidth conservation | Tree structure (top-down) | Medium | Medium | A grid simulator using PARSEC | 2001 | Need to define a threshold for number of requests |
| **Cashing [15]** | A requesting client receives the file and stores a replica of it locally | Response time, Bandwidth conservation | Tree structure (top-down) | Medium | High | A grid simulator using PARSEC | 2001 | _______ |
| **Cascading plus Cashing [15]** | Joining two replication techniques: Cashing and cascading techniques | Response time, Bandwidth conservation | Peer to Peer structure | High | Medium | A grid simulator using PARSEC | 2001 | Need to define a threshold for number of requests |
| **Fast Spread [15]** | If a client requests a file then a replica of file stores at each node along the path toward the client | Response time, Bandwidth conservation | Tree structure (top-down) | Medium | High | A grid simulator using PARSEC | 2001 | Need to storing request history to avoid of double replicating |
| **Least Frequently Used (LFU) [23]** | Always replicates files to local storage, if no space: delete least accessed files | Job execution time | Flat | Low | High | Optorsim | 2003 | Need to files access history |
| **Least Recently Used (LRU) [23]** | Always replicates files to local storage, if no space: delete oldest file in the storage | Job execution time | Flat | Low | High | Optorsim | 2003 | Need to files access history |

TABLE II. COMPARATIVE STUDY ON REPLICATION TECHNIQUES (B)

| Replication technique | Method | Performance metric | Topology | Scalability | Used storage | Simulator | Year | Additional feature |
|---|---|---|---|---|---|---|---|---|
| **Proportional Share Replication (PSR) [24]** | Calculates an ideal workload and distributes replicas | Mean of response time | Tree structure (top-down) | Medium | High | NS2 network simulator (modified) | 2004 | Need to define ideal workload |
| **Bandwidth Hierarchy Replication (BHR) [25]** | Replicates files which are likely to be used frequently within the region in near future | Total job execution time | Hierarchy structure | High | Medium | Optorsim | 2004 | Need to define network-level locality and regions |
| **Simple Bottom-Up (SBU) [16]** | Creates replicas as close as possible to the clients that request the data files with high rates exceeding the pre-defined threshold | Replication frequency, Bandwidth cost, Response time | Tree structure (bottom-up) | Medium | Low | DRepSim (a multi-tier grid simulator) | 2005 | Need to process records in the access history individually |
| **Aggregate Bottom-Up (ABU) [16]** | Aggregates the history records to the upper tier step by step till it reaches the root | Replication frequency, Bandwidth cost, Response time | Tree structure (bottom-up) | Medium | Low | DRepSim (a multi-tier grid simulator) | 2005 | Need to access history |
| **Multi-objective approach [26], [27]** | Reallocates replicas to new candidate sites if a performance metric degrades significantly over best k-time periods | Average response time | Tree structure (top-down) | Medium | Medium | Optorsim | 2006 | Need to calculate replica relocation cost |

| Weight-based replication [28] | Calculates the weight of replica based on the access time in the future time window, based on the last access history | Effective network usage, Mean job execution time | Flat | Low | Medium | Optorsim | 2008 | Need to access history that define based on zip-like distribution |
| --- | --- | --- | --- | --- | --- | --- | --- | --- |
| Least Access Largest Weight (LALW) [29] | Selects a popular file for replication and calculates a suitable number of copies and grid sites for replication | Network usage, Mean job execution time | Hierarchy structure | High | Medium | Optorsim | 2008 | Need to find out a popular file and suitable site |
| Agent based replication [7] | By an agent for each site that holding the master copies, select a candidate site for the placement of replica that exceeds the conditions | Execution time test, Data availability test | Flat | Low | Low | GridSim | 2009 | Need to define agents |

TABLE III. COMPARATIVE STUDY ON REPLICATION TECHNIQUES (C)

| Replication technique | Method | Performance metric | Topology | Scalability | Used storage | Simulator | Year | Additional feature |
| --- | --- | --- | --- | --- | --- | --- | --- | --- |
| Adaptive Popularity Based Replica Placement (APBRP) [30] | Selects a threshold value related to file popularity and places replicas close to clients to reduce data access time while still using network and storage resource efficiency | Storage cost, Average bandwidth cost, Job execution time | Tree structure | Medium | Medium | Optorsim | 2010 | Need to determines threshold value dynamically, based on data request arrival rates |
| Efficient replication strategy [31] | Takes into account the dynamic of sites. Exploits the replicas placement and file request in order to converge towards a global balancing of grid load | Response time, Effective Network Usage | Flat | Low | Medium | Optorsim | 2010 | Need to considering dynamicity of sites |
| Value Based Replication Strategy (VBRS) [32] | Calculates the ideal threshold to decide whether the file should be copied or not. Chooses the replica that should be replaced based on the values of the local replicas | Mean job time, Effective Network Usage | Flat | Low | Low | Optorsim | 2010 | Need to define threshold |
| Enhanced Fast Spread (EFS) [33] | Uses a dynamic threshold that determines if the requested replica should be stored at each node along its path to the requester. Keeps only the important replicas while other less important replicas are replaced with more important replicas | Total response time, Total bandwidth consumption | Flat | Low | Medium | An event-driven simulator written in java | 2011 | Need to frequency of requests, the size of the replica and the last time that the replica was requested |
| Predictive Hierarchical Fast Spread (PHFS) [34] | Tries to forecast future needs and pre-replicates the min hierarchical manner. Uses the hierarchical replication to optimize the utilization of resources | Average access latency | Tree structure | Medium | Medium | Optorsim | 2011 | Need to considering spatial locality and using predictive methods |
| Dynamic Hierarchical Replication (DHR) [35] | Selects best replica when various sites hold replicas. Places replicas in appropriate sites that has the highest number of access for that particular replica | Mean job execution time | Hierarchy structure | High | Low | Optorsim | 2012 | Need to access history |
| Modified Least Access Largest Weight (MLALW) [36] | Stores each replica in an appropriate site. Deletes files by considering least frequently used replicas, least recently used replicas and the size of the replica factors | Effective network usage, Mean job execution time | Hierarchy structure | High | Low | Optorsim | 2012 | Need to LRU lists of replicas, LFU lists of replicas and access history |

## VI. CONCLUSION

Replication is a technique used in grid environments that helps to reduce access latency and network bandwidth utilization. Replication also increases data availability thereby enhancing system reliability. This technique appears clearly applicable to data distribution problems in large scale scientific collaborations, due to their globally distributed user communities and distributed data sites.

In this paper, a review and a comparative study has been done on basic and new replication techniques that have been implemented in grids. After a brief introduction, an overview of grid systems, types of grids and grid topologies were presented in Section 2. In Section 3, replication scenario, challenges and ways of evaluating replication techniques were described. In Section 4, a closer look was taken on twenty two of the various existing data replication strategies. In Section 5, a full comparative study was presented on the replication techniques that were discussed in Section 4. And finally, in this section, a table is presented that shows the results of discussed replication techniques.

Table 4 shows the summary and some results of replication techniques that discussed in Section 5.

TABLE IV. SUMMARIZES THE MAJOR RESULTS OF REPLICATION TECHNIQUES IN GRIDS

| Replication technique | Results and Points |
|---|---|
| Best Client [15] | - Faster average response time than No Replication strategy<br>- Not good overall performance<br>- Not suitable for grid |
| Cascading [15] | - Has an small degree of locality<br>- Not good performance for random access pattern |
| Cashing [15] | - Similar performance as cascading<br>- High response time |
| Cascading plus Cashing [15] | - Client can act as server for sibling<br>- Better performance than cascading<br>- Better performance than cashing |
| Fast Spread [15] | - Consistent performance<br>- High I/O and CPU load<br>- High storage request<br>- Good performance for random access pattern |
| Least Frequently Used (LFU) [23]<br>Least Recently Used (LRU) [23] | - Upgrades overall performance<br>- Upgrades utilization of replica<br>- Better performance than No Replication strategy |
| Proportional Share Replication (PSR) [24] | - Load sharing among replica sites<br>- Better results over cascading technique |
| Bandwidth Hierarchy Replication (BHR) [25] | - Maximizes network-level locality<br>- Good scalability<br>- Better total job times than LRU and LFU |
| Simple Bottom-Up (SBU) [16]<br>Aggregate Bottom-Up (ABU) [16] | - Better results over Fast Spread technique |
| Multi-objective approach [26], [27] | - Good performance in dynamic environments<br>- Dynamic maintainability when performance metric degrades |
| Weight-based replication [28] | - Better performance than LRU and LFU<br>- Has not tested in the real grid systems |
| Least Access Largest Weight (LALW) [29] | - Increases the effective network usage<br>- Better job execution time and effective network usage than LRU, LFU and BHR |
| Agent based replication [7] | - Admissible aggregated response time and data transfer time |
| Adaptive Popularity Based Replica Placement (APBRP) [30] | ▪ Improves access time from the client`s perspective<br>▪ Better performance than Best client, Cascading, Fast Spread, ABU and LRU |
| Efficient replication strategy [31] | ▪ Improves the response time<br>▪ Increases data availability<br>▪ Reduces bandwidth consumption |
| Value Based Replication Strategy (VBRS) [32] | ▪ Decreases network latency<br>▪ Improves performance of the hole system |
| Enhanced Fast Spread (EFS) [33] | ▪ Improves total of response time<br>▪ Improves total bandwidth consumption<br>▪ Enhanced version of Fast Spread for replication strategy in data grid |
| Predictive Hierarchical Fast Spread (PHFS) [34] | ▪ Optimizes the utilization of resources<br>▪ Decreases access latency in multi-tier data grids<br>▪ Improved version of common Fast Spread<br>▪ Lower latency and better performance compared with common Fast Spread |
| Dynamic Hierarchical Replication (DHR) [35] | ▪ Prevents unnecessary creation of replica<br>▪ Efficient storage usage<br>▪ Minimizes access latency |
| Modified Least Access Largest Weight (MLALW) [36] | ▪ Modified version of LALW strategy<br>▪ Better performance than LRU, LFU, BHR, LALW and DHR |

AUTHORS PROFILE

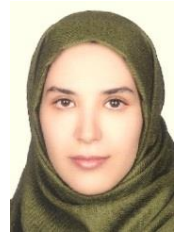

**Sheida Dayyani** was born in Isfahan, Iran 1989. She received the B.Sc. degree in Computer Software Engineering from Sheikh Bahaei University (SHBU), Esfahan, Iran in 2011. Currently, she is a student of Master of Science at Department of Computer Engineering, Sheikh Bahaei University, Esfahan, Iran. Her research areas are Grid Computing, Data Replication and Scheduling algorithms.

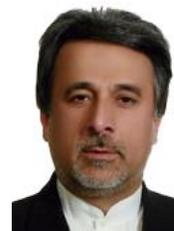

**Mohammad Reza Khayyambashi** was born in Isfahan, Iran in 1961. He received the B.Sc. degree in Computer Hardware Engineering from Tehran University, Tehran, Iran in 1987. He received his M.Sc. in Computer Architecture from Sharif University of Technology (SUT), Tehran, Iran in 1990. He got his Ph.D. in Computer Engineering, Distributed Systems from University of Newcastle upon Tyne, Newcastle upon Tyne, England in 2006. He is now working as a lecturer at the Department of Computer, Faculty of Engineering, University of Isfahan, Isfahan, Iran. His research interests include Distributed Systems, Networking, Fault Tolerance and E-Commerce. He has published in these areas extensively.